\documentclass{INTERSPEECH2023}

\usepackage{microtype}
\usepackage{graphicx}
\usepackage{subfigure}
\usepackage{booktabs} 

\usepackage{amsfonts}
\usepackage{url}

\usepackage{hyperref}

\usepackage{algorithm}
\usepackage{algorithmic}

\usepackage[textsize=tiny]{todonotes}

\title{Style Description based Text-to-Speech with  Conditional Prosodic Layer Normalization based Diffusion GAN }
\name{Neeraj Kumar$^1$, Ankur Narang$^2$, Brejesh Lall$^1$}
\address{
  $^1$ IIT Delhi\\
  $^2$ IEEE Senior Member}
\email{neerajkr2k14@gmail.com, annarang2@gmail.com, brejesh@ee.iitd.ac.in}

\begin{document}

\maketitle

\begin{abstract}
In this paper, we present a Diffusion GAN based approach (Prosodic Diff-TTS) to generate the corresponding high-fidelity speech based on the style description and content text as an input to generate speech samples within only 4 denoising steps. It leverages the novel conditional prosodic layer normalization to incorporate the style embeddings into the multi head attention based phoneme encoder and mel spectrogram decoder based generator architecture to generate the speech. The style embedding is generated by fine tuning the pretrained BERT model on auxiliary tasks such as pitch, speaking speed, emotion,gender classifications. We demonstrate the efficacy of our proposed architecture on multi-speaker LibriTTS and PromptSpeech datasets, using multiple quantitative metrics that measure generated accuracy and MOS.
\end{abstract}

\vspace{-1.0em}

\section{Introduction}

Text to speech (TTS) tries to synthesise a genuine and comprehensible voice from text and garners significant interest from the machine learning field \cite{Arik2017DeepVR,Ren2019FastSpeechFR,Shen2017NaturalTS}. TTS models are able to synthesis natural human speech when trained with a huge quantity of high-quality and single-speaker recordings and this capability has been extended to multi-speaker settings \cite{Ping2017DeepV3,Kumar2021NormalizationDZ,Sun2019TokenLevelED}. Today, custom voice is gaining popularity in a variety of application situations, including personal assistant, news broadcast, and audio navigation, and is extensively supported by commercial speech platforms. 

Prior research on TTS has focused on regulating certain style aspects, including prosody control using word-level prosody tags\cite{Guo2022UnsupervisedWP}, speaking speed control\cite{Bae2020SpeakingSC} using sentence-level speaking-rate, and pitch control using pitch contours. All previous works require users to input the precise style factor value with acoustic expertise or choose the reference speech that matches the requirements, which is time-consuming and not user-friendly.  There is a trade off between fine-tuning parameters and voice quality while converting the source TTS model to a new voice which is frequently recorded in various speaking styles, emotions, dialects, and surroundings. Therefore, style control with natural language text is preferable.

We investigate using a text description(prompt) to guide speech synthesis. The input prompt consists of a style description and a content description with a colon in between. “A lady whispers to her friend slowly: everything will go OK, right?” requires the model to synthesis speech with the content “everything will go fine, right?” in a female voice, a slow speaking tempo, and a whispering manner. Users can compose speech from a style text without acoustic knowledge or reference speech, allowing style freedom.

Some of the existing work, PromptTTS\cite{Guo2022PromptTTSCT}has used style encoder to extract the style token from style description and content encoder and speech decoder to generate the final speech. We have proposed the diffusion model based framework which has the ability to model complex data distribution to solve a variety of speech synthesis problems\cite{Huang2022FastDiffAF,Liu2022DiffGANTTSHA}. We have used the denosing diffusion GAN\cite{Xiao2021TacklingTG} that uses transformer\cite{Vaswani2017AttentionIA} based encoder-decoder based generator architecture to generate the mel-spectrogram conditioned on timesteps, intermediate mel-spectrogram and style embeddings. We have extracted the style tokens from fine tuning the pretrained BERT\cite{Devlin2019BERTPO} on auxiliary task and fed into the denosing diffusion GAN through the proposed conditional prosodic layer normalization.
\vspace{-0.25em}
Our contribution are as follows:
\vspace{-0.5em}
\begin{itemize}
\item Partially inspired by the denoising diffusion GAN\cite{Xiao2021TacklingTG}, We model the denoising distribution using a conditional generator that has been adversarially trained to match the actual denoising distribution. Prosodic Diff-TTS permits larger denoising steps at inference, hence drastically reducing the number of denoising steps and accelerating sampling. 
\item We have used the pretrained BERT model to learn the 128 dimensional style token in multi task learning fashion using the cross entropy loss for optimization.
\item To make the generator conditional on style tokens, we have proposed Conditional Prosodic Layer normalization to inject the style into the denoising generator model to learn the given style information as well as other style-agnostic prosodic variations.
\item Using extensive experiments on multi-speaker PromptSpeech \cite{Guo2022PromptTTSCT} and LibriTTS\cite{Panayotov2015LibrispeechAA} datasets, we show both qualitative and quantitative results along with high-quality output speech given the input style text and content description.
\end{itemize}

\vspace{-1em}
\section{Model Architecture}
\vspace{-0.5em}
Commonly, diffusion models assume the denoising distribution can be approximated by Gaussian distributions, necessitating a lengthy reverse operation. As the denoising step is increased and the non-Gaussian data distribution is present, the underlying denoising distribution becomes more complex and multimodal. We have adopted the denosing diffusion GAN to model the multimodal denoising distribution where the generator produces the  mel-spectrogram($x_{0}$) given the content($y$) and style token through the conditional prosodic layer normalization. The discriminator distinguishes the fake and real mel-spectrograms conditional on style embeddings.

\begin{figure}[h]
  \begin{center}
   \includegraphics[width=0.8\linewidth]{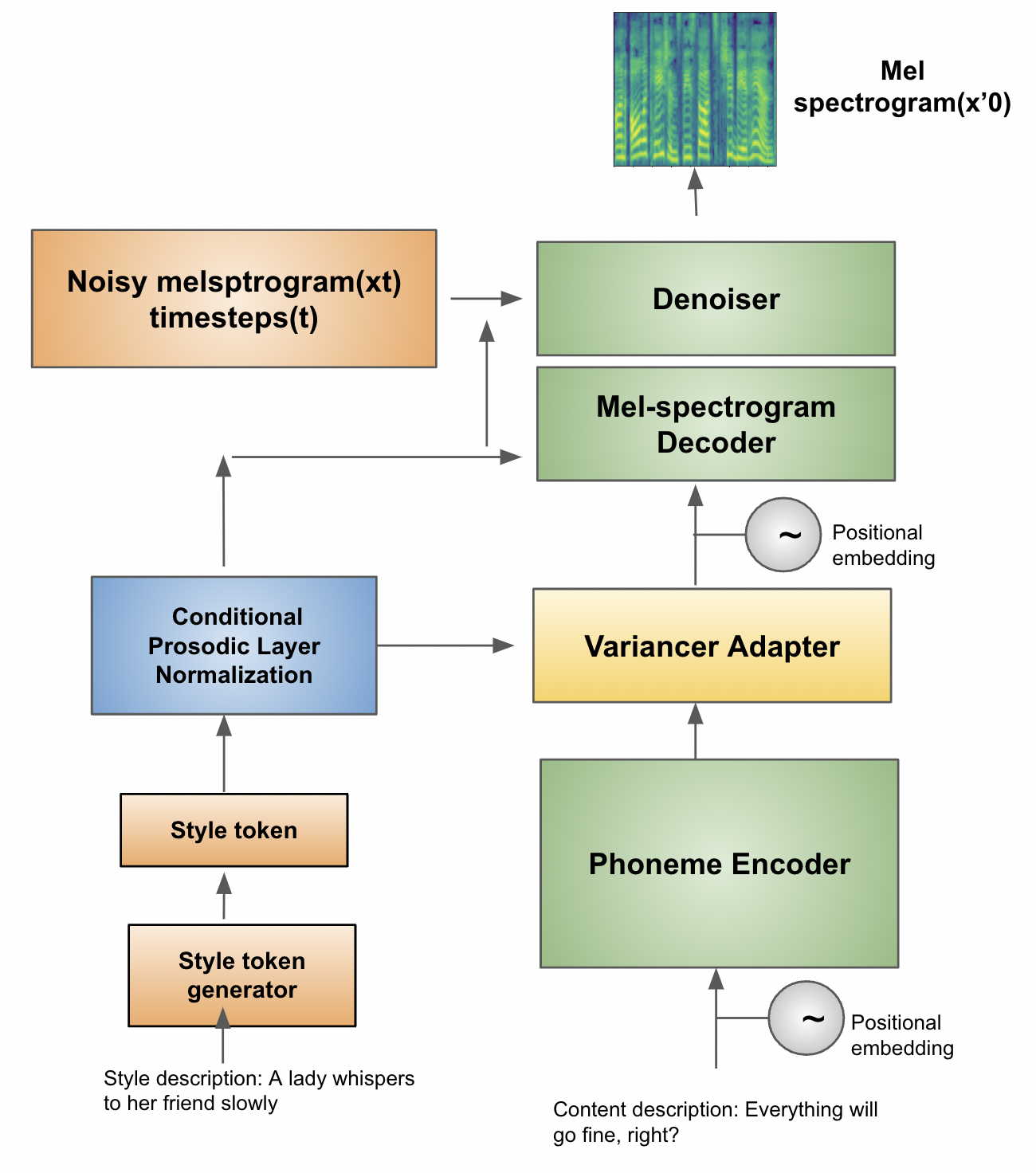}\vspace{-1em}
   \caption{Prosodic Diff-TTS architecture}
   \label{fig:prod}
  \end{center}\vspace{-1.5em}
\end{figure}

\begin{figure}[h]
  \begin{center}
   \includegraphics[width=0.8\linewidth]{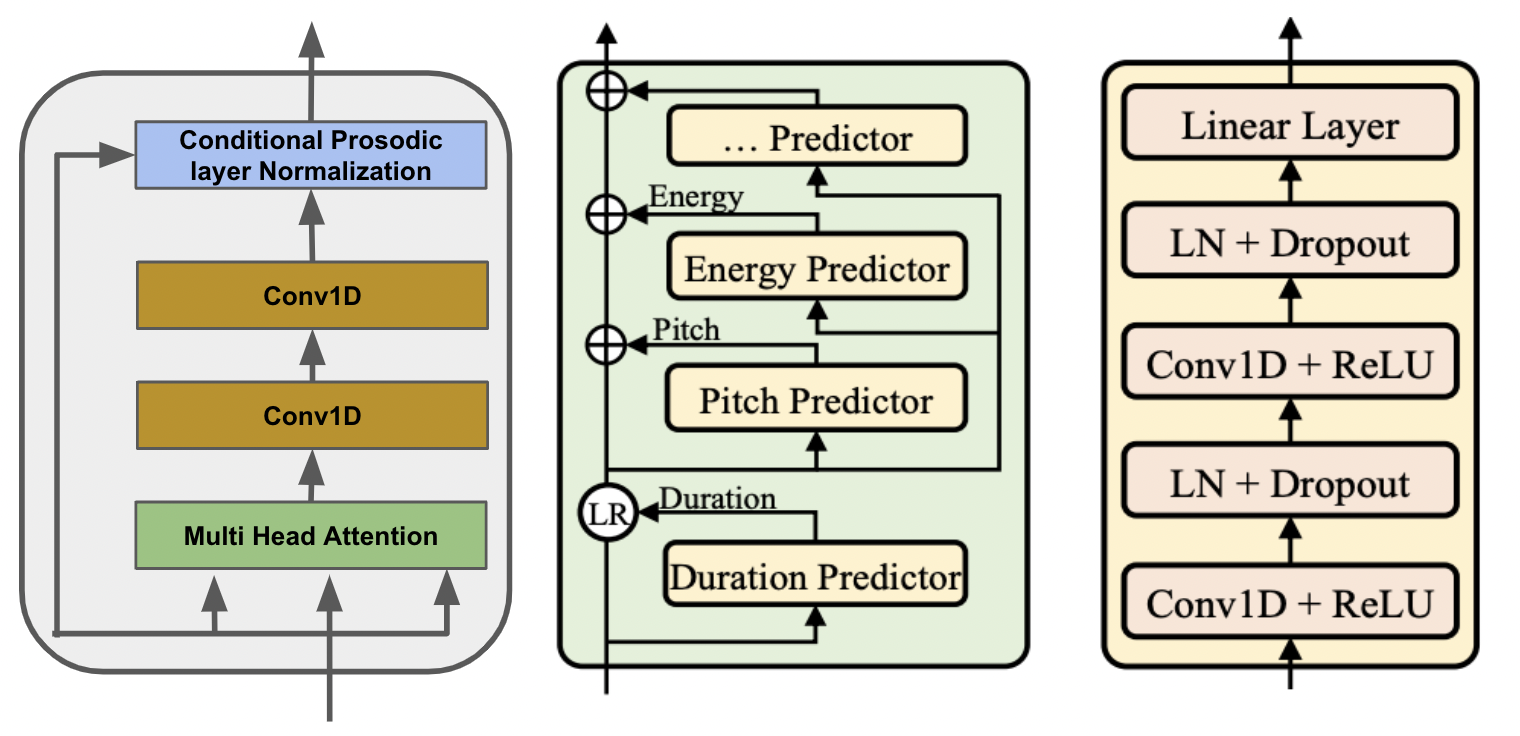}\vspace{-1em}
   \caption{Left: Multi Head Attention block , Centre : Variance Adapter\cite{Ren2020FastSpeech2F} , Right
: Variance Predictor}
   \label{fig:varpred}
  \end{center}\vspace{-2em}
\end{figure}
\vspace{-1em}
\subsection{Generator}
As illustrated in Figure~\ref{fig:prod}, Prosodic Diff-TTS takes phoneme sequence (denoted as $\mathbf{y}$) as input to generate  mel-spectrogram features $\mathbf{x}'_0$ with generator and then uses a HiFi-GAN-based neural vocoder \cite{Kong2020HiFiGANGA} to produce time-domain waveforms. The generator architecture uses the phoneme encoder, conditional style layer normalization framework, variance adaptor, mel-spectrogram decoder along with denoiser conditioned on style token, timesteps and noisy melspectrogram to generate mel-spectrogram features $\mathbf{x}_0$. The phoneme encoder and mel-spectrogram decoder is based on a multi-head self-attention network, and position feed-forward network which consists of two Conv1D and normalization stages. The proposed method stacks multiple multi-head self-attention\cite{Vaswani2017AttentionIA} blocks with phoneme embedding and position encoding as an input at the encoder side(Fig. 1), and multiple multi-head self-attention blocks with position encoding,style token and output from variance adapter\cite{Ren2020FastSpeech2F}(Fig. \ref{fig:varpred}) for the  mel-spectrogram generation at the decoder side. 

\begin{figure}[h]
  \begin{center}
   \includegraphics[width=0.8\linewidth]{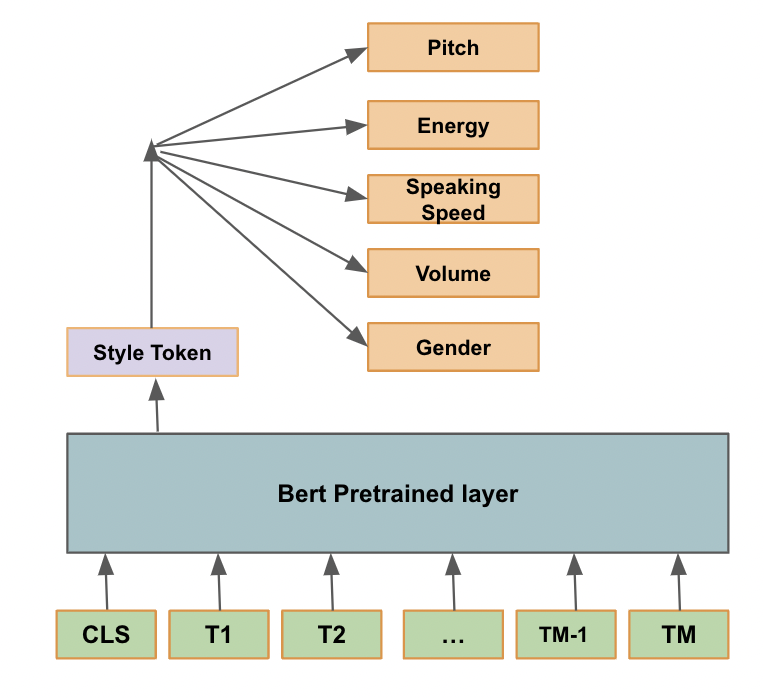}\vspace{-1em}
   \caption{Style token Generator using pretrained BERT }
   \label{fig:beretp}
  \end{center}\vspace{-1em}
\end{figure}

\begin{figure}[h]
  \begin{center}
   \includegraphics[width=1.0\linewidth]{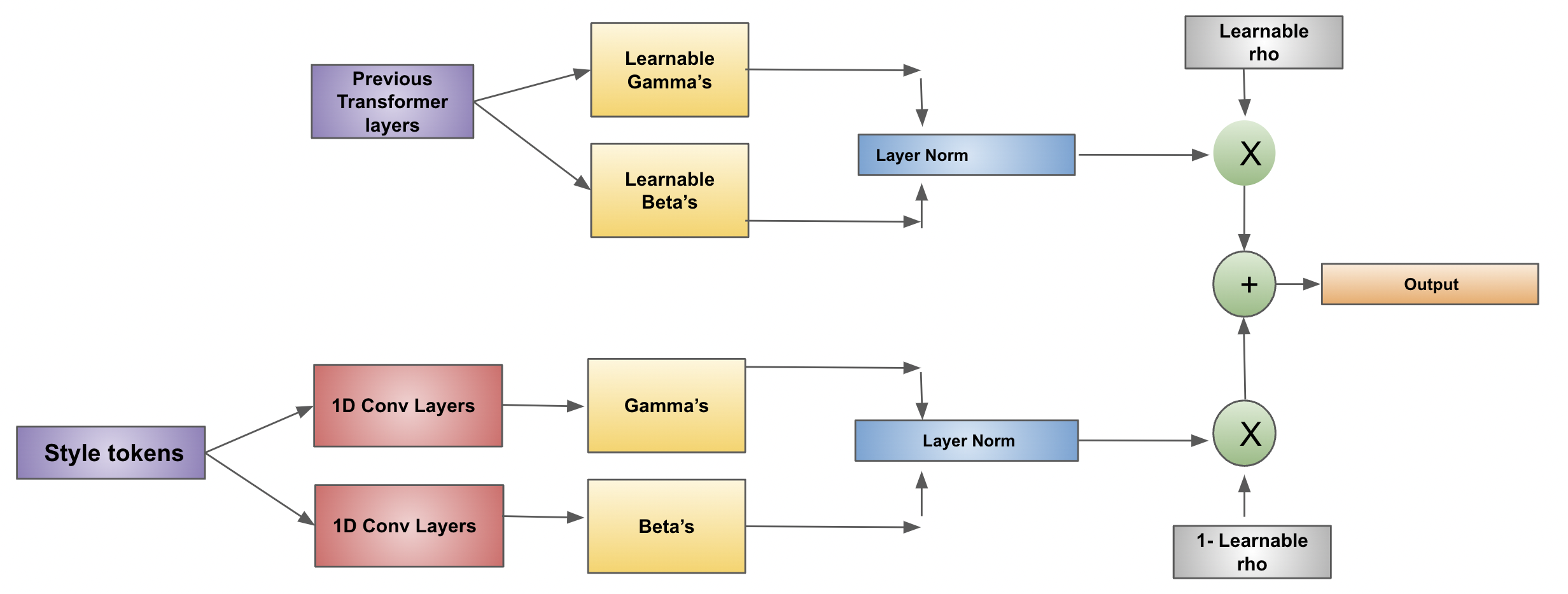}\vspace{-1em}
   \caption{Conditional Prosodic Layer Normalization architecture}
   \label{fig:normcond}
  \end{center}\vspace{-2em}
\end{figure}

\vspace{-1em}
\subsection{Style token generator}
\vspace{-0.5em}
We have trained the pretrained BERT to generate the 128 dimensional style embeddings in multi task learning fashion with auxiliary tasks related to prosodic features from the style text such as pitch , gender, speaking speed, emotion and volume as shown in Figure \ref{fig:beretp}.The input style text sequence T = [T1, T2, $\cdots$ , TM] is prepended with a [CLS] token, converted into a word embedding, and fed into the BERT model, where M refers to the length of style text.The hidden vector corresponding to the [CLS] token is regarded as the style representation to guide the content encoder and the speech decoder.
\vspace{-1em}
\subsection{Proportional Prosodic Layer Normalization}
\vspace{-0.5em}

Layer normalization could significantly impact the hidden activation and final prediction through learnable scale and bias in multi head attention block as given in equations \ref{eq:2} and \ref{eq:3}. We have proposed a conditional prosodic style layer normalization(Fig. \ref{fig:normcond}) which is employed at the phoneme encoder, mel-spectrogram decoder and denoiser module. The style embeddings are passed into the linear layer to generate the style related affine parameters namely $\gamma_{style}$ and $\beta{style}$. The affine parameters of layer normalization will learn the style agnostic features. Both are combined by equation \ref{eq:4} and $\rho$ is used to control the amount of information flow through normalization.The value of $\rho$ is constrained to the range of [0, 1] simply by imposing bounds at the parameter update step. This normalization will help the variance predictors to predict the duration, energy and fundamental frequency such that it incorporates the style of the input style text and content of the input content text.

\vspace{-0.5em}
\begin{equation}
\label{eq:2}
    \mu\textsubscript{i} = \frac{1}{m}\sum_{j=1}^{m} x_{ij}, 
\sigma\textsubscript{i}^2 = \frac{1}{m}\sum_{j=1}^{m}(x_{ij} - \mu\textsubscript{i})^2
\end{equation}
\vspace{-0.5em}
\begin{equation}
\label{eq:3}
\hat{x}_{i,j} = \frac{x_{ij} - \mu\textsubscript{i}}{\sqrt{\sigma_{i}^2}}
\end{equation}
\vspace{-0.5em}
\begin{equation}
\label{eq:4}
    \hat{y}_{ij} = \rho(\gamma\textsubscript{LN} \hat{x}_{ij} + \beta\textsubscript{LN}) + (1- \rho)(\gamma\textsubscript{style} \hat{x}_{ij} + \beta\textsubscript{style})
\end{equation}

\begin{figure}[h]
  \begin{center}
   \includegraphics[width=0.8\linewidth]{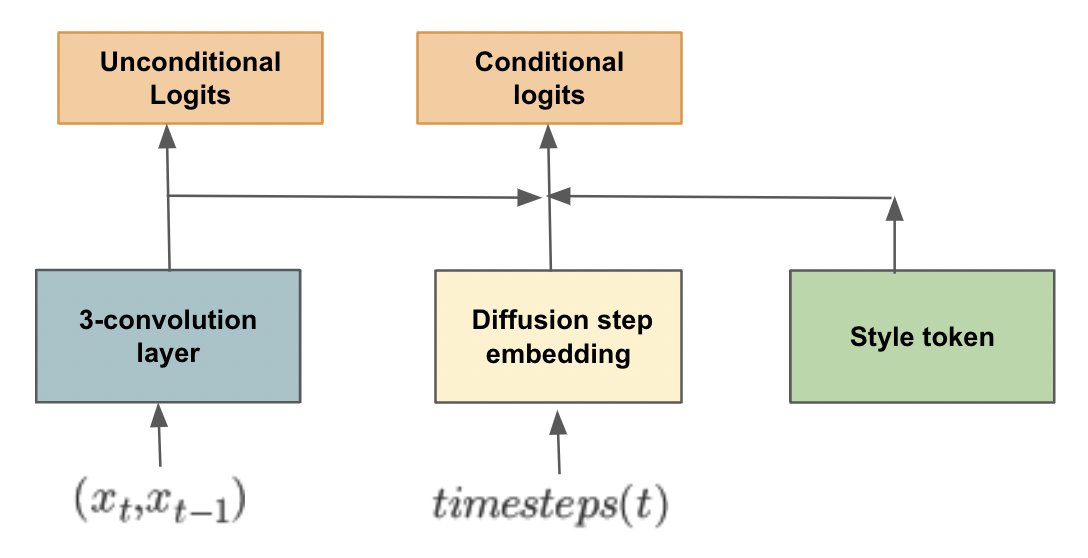}\vspace{-1em}
   \caption{Discriminator architecure of Prosodic Diff-TTS}
   \label{fig:discf}
  \end{center}\vspace{-3em}
\end{figure}

\subsection{Discriminator}
The discriminator\cite{Yang2020VocGANAH} is used to distinguish the fake and real melspectrogram($x_{t},x_{t-1}$) using unconditional and conditional logits which is conditioned on time-steps and style embeddings.We have used 1D convolution based network with leakyRelu as activation function to predict the unconditional logits and conditional logits based on style tokens as shown in Figure \ref{fig:discf}.

\vspace{-0.5em}
\subsection{Training Loss}
\vspace{-0.5em}
We focus on discrete-time diffusion models, where denoising steps are large, and use a conditional GAN to model the denoising distribution. Prosodic Diff-GAN trains a conditional GAN-based generator $p_\theta(\mathbf{x}_{t-1}|\mathbf{x}_t)$ to approximate the true denoising distribution $q(\mathbf{x}_{t-1}|\mathbf{x}_t)$ with an adversarial loss that minimizes a divergence $D_{\text{adv}}$ per denoising step:
\vspace{-0.5em}
\begin{equation}
    \min_{\theta}\sum_{t\geq 1}\mathbb{E}_{q(\mathbf{x}_t)}[D_{\text{adv}}(q(\mathbf{x}_{t-1}|\mathbf{x}_t)||p_\theta(\mathbf{x}_{t-1}|\mathbf{x_t}))],
\end{equation}
where we adopt the least-squares GAN (LS-GAN) training formulation \cite{Mao2016LeastSG} to minimize $D_{\text{adv}}$

The discriminator is trained to minimize the loss
\vspace{-0.5em}
\begin{equation} \label{eq:d-loss}
\begin{split}
    \mathcal{L}_D = \sum_{t\geq 1}\mathbb{E}_{q(\mathbf{x}_t)q(\mathbf{x}_{t-1}|\mathbf{x}_t)}[(D_\phi(\mathbf{x}_{t-1}, \mathbf{x}_t, t, s)-1)^2] \\ + \mathbb{E}_{p_\theta(\mathbf{x}_{t-1}|\mathbf{x}_t)}[D_\phi(\mathbf{x}_{t-1}, \mathbf{x}_t, t, s)^2].
\end{split}
\end{equation}
\vspace{-0.5em}
The generator is trained adversarially($\mathcal{L}_{adv}$) to minimize the equation, so that it could generate a realistic  mel-spectrogram. The variance predictor, namely the duration($\mathcal{L}_{duration}$), energy($\mathcal{L}_{energy}$ and  pitch($\mathcal{L}_{pitch}$ use MSE loss to optimize their network.  The mean absolute error is also used on mel-spectrogram to optimize generator. 
\vspace{-0.5em}
\begin{multline} \label{eq:g-loss}
    \mathcal{L}_G = \mathcal{L}_{adv} + \lambda_{duration}\mathcal{L}_{duration} +\lambda_{energy}\mathcal{L}_{energy}+ \\ 
    \lambda_{pitch}\mathcal{L}_{pitch}+ \lambda_{fm}\mathcal{L}_{fm},
\end{multline}
where $\lambda_{duration}$,$\lambda_{energy}$,$\lambda_{pitch}$,$\lambda_{fm}$ are the training hyperparameters.
\vspace{-0.5em}
\begin{equation}
    \mathcal{L}_{adv}=\sum_{t\geq 1}\mathbb{E}_{q(\mathbf{x}_t)}\mathbb{E}_{p_\theta(\mathbf{x}_{t-1}|\mathbf{x}_t)}[(D_\phi(\mathbf{x}_{t-1}, \mathbf{x}_t, t, s)-1)^2],
\end{equation}

To avoid the mode collapse, the feature matching loss\cite{Larsen2015AutoencodingBP}  $\mathcal{L}_{fm}$, is used on generator by summing l1 distances between every discriminator feature maps of real and generated samples:
\vspace{-0.5em}
\begin{equation}
\resizebox{0.48\textwidth}{!}{
    $\mathcal{L}_{fm} = \mathbb{E}_{q(\mathbf{x}_t)}[\sum_{i=1}^N||D_\phi^i(\mathbf{x}_{t-1}, \mathbf{x}_t, t, s)-D_\phi^i(\mathbf{x}'_{t-1}, \mathbf{x}_t, t, s)||_1]$,
}
\end{equation}
\vspace{-0.5em}
\subsection{Training and Inference algorithm of Prosodic Diff-TTS }

\begin{algorithm}[h]
  \caption{Training procedure of Prosodic Diff-TTS}
  \label{alg:1}
\textbf{Input}: The generator $G_\theta$ with parameters $\theta$. The discriminator $D_\phi$ with parameter $\phi$. A pre-calculated variance schedule $\beta_1, \cdots,\beta_T$ with $T$ diffusion steps with style token s.
    \begin{algorithmic}[1]
    \REPEAT 
    \STATE Sample $\mathbf{x}_0 \sim q(\mathbf{x})$, $\epsilon\sim\mathcal{N}(\mathbf{0}, \mathbf{I})$, and $t\sim\mathrm{Unif}(\{1,\cdots,T\})$
    \STATE Sample $\mathbf{x}_t$ and $\mathbf{x}_{t-1}$ given $\mathbf{x}_0$ according to $q(\mathbf{x}_{1:T}|\mathbf{x}_0)=\prod_{t\geq 1}q(\mathbf{x}_t|\mathbf{x}_{t-1}), \quad q(\mathbf{x}_t|\mathbf{x}_{t-1}):=\mathcal{N}(\mathbf{x}_t; \sqrt{1-\beta_t}\mathbf{x}_{t-1},\beta_t\mathbf{I})$ 

    \STATE $\mathbf{x}'_{0}$ = $G_\theta(x_{t},y,s,t)$
    \STATE Sample $\mathbf{x}'_{t-1} \sim$ $q(\mathbf{x}_{t-1}|\mathbf{x}_{t},\mathbf{x}'_{0})$
    \STATE Do back-propagation with $\mathcal{L}_G$ and update $\theta$ one step with gradient descent;
    \STATE Do back-propagation with $\mathcal{L}_D$ and update $\phi$ one step with gradient descent;
    \UNTIL{refinement model  converged}
  
    \end{algorithmic}
    \end{algorithm}

 \begin{algorithm}[h]
  \caption{Inference procedure of Prosodic Diff-TTS}
  \label{alg:2}
  \textbf{Input}: A trained generator $G_\theta$ and one testing sample $(\mathbf{y}, s)$.
\begin{algorithmic}[1]
  \STATE Sample $\mathbf{x}_T\sim\mathcal{N}(\mathbf{0}, \mathbf{I})$;
  \STATE $\mathbf{x}_t \gets \mathbf{x}_T$;
  \FOR{$t=T,T-1,...,1$}
    \STATE Forward-propagate $(\mathbf{x}_t, \mathbf{y}, s, t)$ to $G_\theta$ to calculate $\mathbf{x}'_0$;
    \STATE Sample $\mathbf{x}'_{t-1}$ given $\mathbf{x}'_0$ and $\mathbf{x}_t$ by $q(\mathbf{x}_{t-1}|\mathbf{x}_{t},\mathbf{x}'_{0})$;
    \STATE $\mathbf{x}_t \gets \mathbf{x}'_{t-1}$;
  \ENDFOR
  \STATE {\bfseries Return} $\mathbf{x}_t$;
\end{algorithmic}
\end{algorithm} 

\vspace{-2.0em}
\section{Experiments}
\vspace{-1.0em}
\subsection{Datasets} 
\vspace{-0.5em}
We train and evaluate the model on two datasets namely PromptSpeech\cite{Guo2022PromptTTSCT} and LibriTTS\cite{Panayotov2015LibrispeechAA}. PromptSpeech has  5 different style factors (gender, pitch, speaking speed, volume, and emotion) and we extracted the audio from the commercial TTS API \footnote{\url{https://azure.microsoft.com/en-us/services/cognitive-services/text-to-speech/\#overview}}. LibriTTS has 4 different styles (gender, pitch, speaking speed and volume).  The number of training and test samples are 1.5 lakh and 5k respectively for PromptSpeech and 26k and 1.3k respectively for LibriTTS.
\vspace{-0.5em}
\subsection{Training and Preprocesing Steps} We convert the text sequence into the phoneme sequence\cite{DeepSpeech2, tacotron} using open-source grapheme-to phoneme tool\cite{g2p}. We extract the phoneme duration with MFA\cite{mfa}, an open-source system for speech-text alignment to improve the alignment accuracy. We extracted the pitch contour, F0 using PyWorldVocoder tool \footnote{\url{https://github.com/JeremyCCHsu/Python-Wrapper-for-World-Vocoder}} We transfer the raw waveform into melspectrograms by setting the frame size and hop size to 1024 and 256 with respect to the sample rate of 22050 Hz. We have used pretrained HiFi universal vocoder to generate the audio waveforms. 
\vspace{-0.5em}
\subsection{Model Configuration}
We employ a pre-trained BERT model with 12 hidden layers and 110M parameters. On the basis of an auxiliary classification task involving five style parameters, the BERT model is fine tuned.
In the generator architecture's phoneme encoding and output mel-spectrogram decoding stages, four feed forward transformer blocks were utilised. The hidden size, number of attention heads, kernel size, and filter size are set to 256, 2, 9, and 1024, respectively, for the one-dimensional convolution in the multi head attention block. The number of attention heads is set to 2. The denoiser module has 20 residual blocks along with hidden dimension of 512 and dropout set to 0.2.
Two blocks of Conv1D, ReLU, layer normalization, and a dropout layer compose the Variance predictor. The kernel sizes of the 1D-convolution are set to 3, the input/output sizes for both layers are 256/256, and the dropout rate is set to 0.5. The mel-spectrogram that is generated is optimised with mean square error loss.
The network topology of the discriminator with unconditional block and the discriminator with conditional block consists of two 1D convolutional layers. The convolution channels consist of 64, 128, 512, 128, and 1. The kernel sizes are 3, 5, 5, 5, 3, and the strides are 1, 2, 2, 1.

\subsection{Style transfer evaluation on synthesized samples}
\label{sssec:main results}
\vspace{-0.5em}
We have performed the experiment to check whether the input style has been successfully been transferred to the generated speech. We have used the PyWorldVocoder tool to compute the accuracy of synthesized speech on various style tasks such as gender, pitch,speaking speed, volume a classification. We have trained a neural network classifier for emotion classification with more than 98$\% $accuracy. Table~\ref{main result} shows the comparison of Prosodic Diff-TTS with the prior work, PromptTTSwhich shows better performance of the proposed method in most style related task which is attributed to conditional prosodic layer normalization as well as diffusion GAN based architecture which generate high fidelity speech.
\vspace{-1em}
\begin{table}[!h]\footnotesize
    \caption{The accuracy (\%) of PromptTTS and Prosodic Diff-TTS on 1-PromptSpeech and 2-LibriTTS datasets.}
    \label{main result}
    \centering
    \setlength{\tabcolsep}{0.65mm}{
        \begin{tabular}{ccccccc}
            \toprule
            \textbf{Model} &\textbf{Gender} & \textbf{Pitch} & \textbf{Speed}  & \textbf{Volume} & \textbf{Emotion}  & \textbf{Mean} \\
            \midrule
            \midrule
            PromptTTS-1 & \textbf{99.57} & 82.60 & 86.55 & 87.35 & 91.78 & 89.57 \\
            Prosodic Diff-TTS-1 & 99.55 & \textbf{85.72} & \textbf{87.49} & \textbf{88.40} & \textbf{93.45} & \textbf{90.62} \\
            
            \midrule
            PromptTTS-2 & 99.16 & 82.69 & 92.57 & \textbf{89.73} & - & 91.04 \\
            Prosodic Diff-TTS-2 & \textbf{99.28} & \textbf{84.17} & \textbf{93.32} & 88.91 & - & \textbf{92.81} \\
            \bottomrule 
        \end{tabular}
    }
    \vskip -0.16in
\end{table}

\subsection{Speech Quality}
\label{sssec:speech quality}

Twenty samples of speakers are taken for PromptSpeech and LibriTTS test set are used to perform MOS~\cite{Chu2001AnOM} to evaluate the generated samples in terms of naturalness(how the synthesized voices sound natural like human), similarity(how the synthesized voices sound with respect to the input style description).We compare the MOS of audio samples including: (1) GT, the ground-truth recordings; (2) GT mel + HiFiGAN, where we first convert ground-truth speech into mel-spectrogram, and then convert the mel-spectrogram back to speech using HiFiGAN \cite{Kong2020HiFiGANGA}; (3) PromptTTS; (4) Prosodic Diff-TTS. Both systems in (3) and (4) use HiFiGAN as vocoder.  According to Table~\ref{speechquality}, it can be seen that Prosodic Diff-TTS outperforms the PromptTTS slightly in terms of speech quality, as the proposed Prosodic Diff-TTS is able to model the multi modal distribution more efficiently than PromptTTS.Synthesized audio samples are present at this site \footnote{\url{https://sites.google.com/view/prosdifftts}}

\begin{table}[!h]\footnotesize
    \newcommand{\tabincell}[2]{\begin{tabular}{@{}#1@{}}#2\end{tabular}}
    \caption{The results of speech quality with 95\% confidence intervals.}
    \label{speechquality}
    \centering
    \setlength{\tabcolsep}{2.04mm}{
        \begin{tabular}{ccc}
            \toprule
             \textbf{Version} & \textbf{Setting} & \textbf{MOS} \\
            \midrule
            \midrule
            PromptSpeech & GT & 4.15 $\pm$ 0.08  \\
            ~ & GT mel + HiFiGAN & 4.10 $\pm$ 0.08 \\
            ~ & PromptTTS & 3.98 $\pm$ 0.06 \\
            ~ & Prosdic Diff-TTS & 4.01 $\pm$ 0.02  \\
            \midrule
            LibriTTS & GT & 4.21 $\pm$ 0.08  \\
            ~ & GT mel + HiFiGAN & 4.13 $\pm$ 0.07  \\
            ~ & PromptTTS & 3.82 $\pm$ 0.08  \\
            ~ & Prosdic Diff-TTS & 3.85 $\pm$ 0.08  \\
            \bottomrule 
        \end{tabular}
    }
    \vskip -0.16in
\end{table}
\vspace{-0.5em}
\subsection{Qualitative Results}
\begin{figure}[h]
  \begin{center}
   \includegraphics[width=1.0\linewidth]{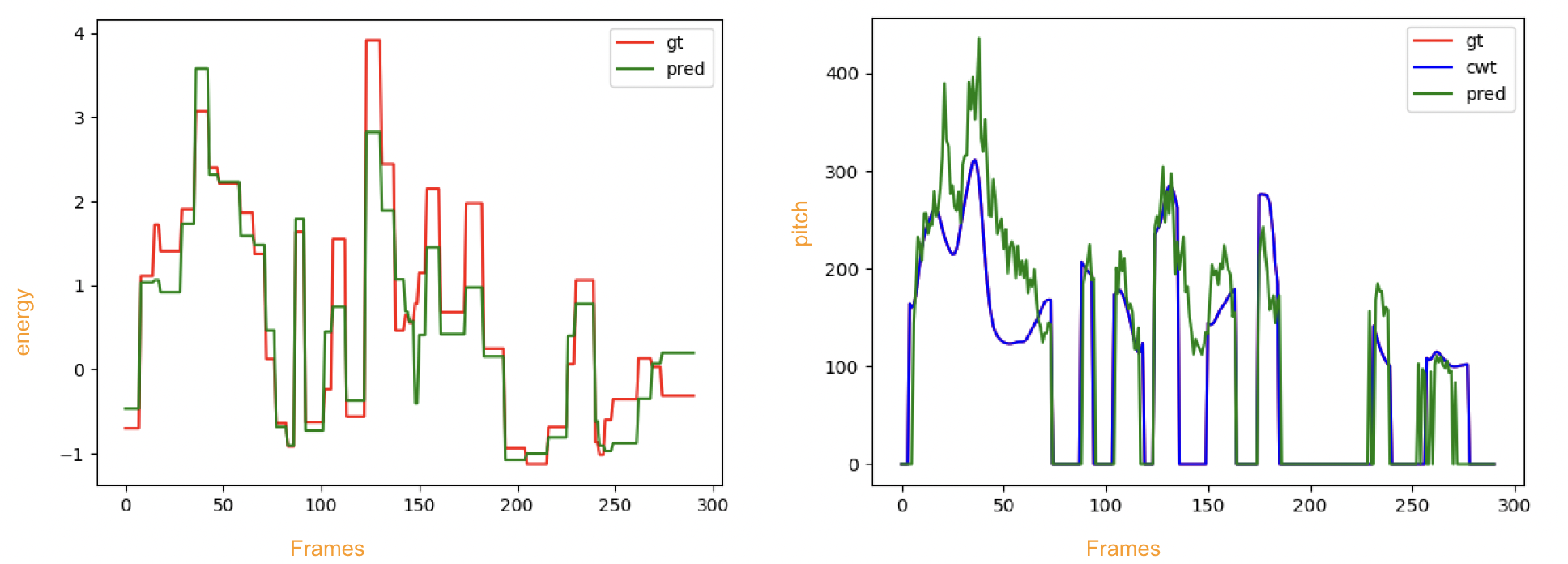}\vspace{-1em}
   \caption{Plot of energy and pitch of synthesized samples with real speech(Content: Your memory must be conveniently short,"" chafed the master , Style: Please say a loud girl with a bass)}
   \label{fig:ptichenergy}
  \end{center}\vspace{-3em}
\end{figure}

\begin{figure}[h]
  \begin{center}
   \includegraphics[width=0.8\linewidth]{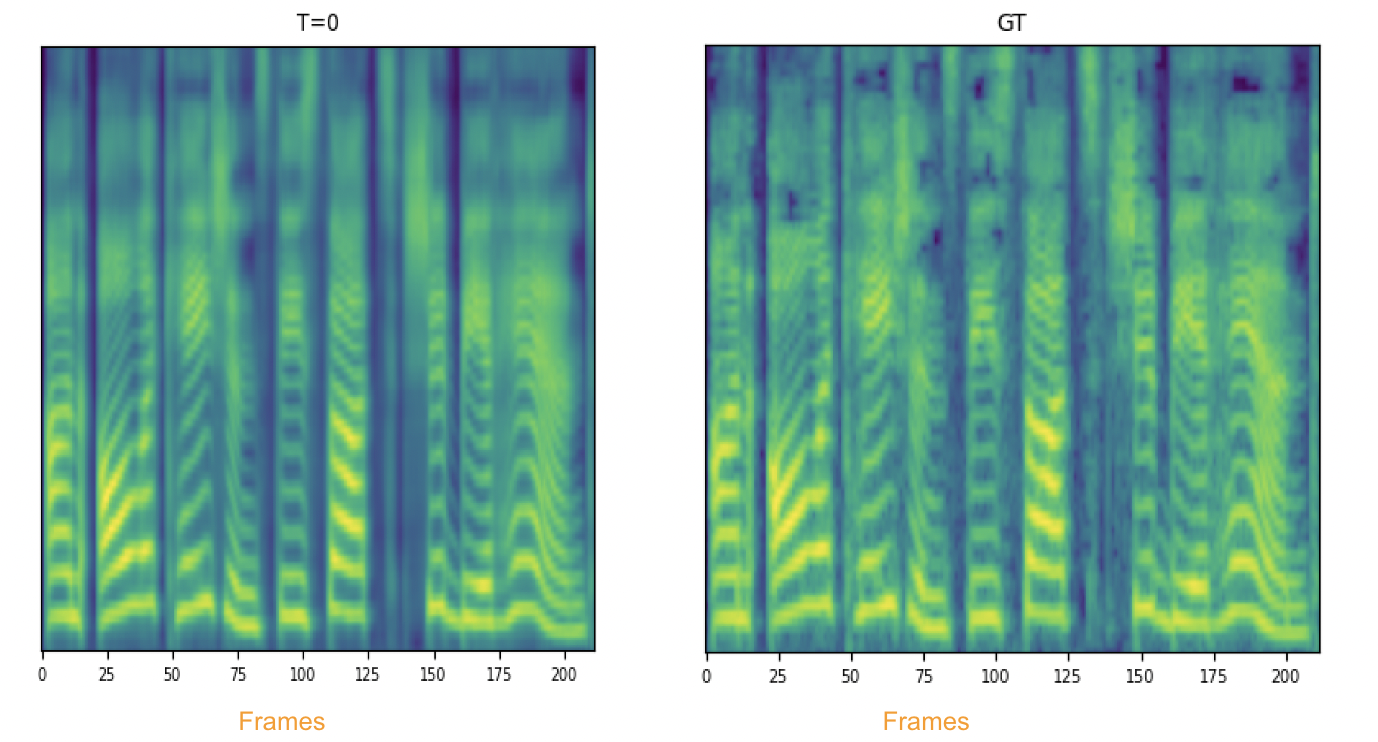}\vspace{-1em}
   \caption{Plot of mel-spectrogram of synthesized samples with real speech.(Content: Your memory must be conveniently short,"" chafed the master , Style: Please say a loud girl with a bass)}
   \label{fig:melgt}
  \end{center}\vspace{-2em}
\end{figure}

\vspace{-0.5em}
We extracted the pitch and energy from the predicted speech and the ground truth speech using the tool. Figure \ref{fig:ptichenergy} shows the prosody of generated samples are similar to that of ground truth samples. Figure \ref{fig:melgt} shows the similarity of predicted and ground truth mel-spectrogram at T=4.

\vspace{-0.5em}
\subsection{Ablation Study}
\vspace{-0.5em}
We have done the ablation study by changing the time-steps with values from T=1,2,4. Table \ref{tab:speechablationdf} shows the better MOS at T=4 as compared to other time-steps. The possible reasons include the difficulty of directly generating samples from a complex distribution in one or two timesteps, and the problem of overfitting when the discriminator only examines clean samples.
\vspace{-1em}
\begin{table}[!h]\footnotesize
    \newcommand{\tabincell}[2]{\begin{tabular}{@{}#1@{}}#2\end{tabular}}
    \caption{The results of speech quality with 95\% confidence intervals at different timesteps}
    \label{tab:speechablationdf}
    \centering
    \setlength{\tabcolsep}{2.04mm}{
        \begin{tabular}{ccc}
            \toprule
             \textbf{Version} & \textbf{Setting} & \textbf{MOS} \\
            \midrule
            \midrule
            PromptSpeech & T=1 & 3.81 $\pm$ 0.09  \\
            ~ & T=2 & 3.95 $\pm$ 0.06 \\
            ~ & T=4 & 4.01 $\pm$ 0.08 \\
            \midrule
            LibriTTS & T=1 & 3.72 $\pm$ 0.10  \\
            ~ & T=2 & 3.79 $\pm$ 0.07  \\
            ~ & T=4 & 3.85 $\pm$ 0.08 \\
            \bottomrule 
        \end{tabular}
    }
    \vskip -0.16in
\end{table}

\vspace{-0.5em}
\section{Conclusion}

In this paper, we have proposed the diffusion GAN based speech synthesis architecture which can generate realistic speech based on input content and style text description. We have proposed the conditional prosodic layer normalization which injects the style into the encoder and decoder of generator architecture at multiple layers through the affine parameters of normalization. We have extracted the 128 dimensional style embedding by fine tuning the pretrained BERT model on multiple auxiliary tasks such as pitch, gender, volume, speaking speed and emotions. Using extensive experiments on multi-speaker datasets(PromptSpeech and LibriTTS), we have shown both qualitative and quantitative results along with high-quality of audio output.


\begin{thebibliography}{10}
\providecommand{\url}[1]{#1}
\csname url@samestyle\endcsname
\providecommand{\newblock}{\relax}
\providecommand{\bibinfo}[2]{#2}
\providecommand{\BIBentrySTDinterwordspacing}{\spaceskip=0pt\relax}
\providecommand{\BIBentryALTinterwordstretchfactor}{4}
\providecommand{\BIBentryALTinterwordspacing}{\spaceskip=\fontdimen2\font plus
\BIBentryALTinterwordstretchfactor\fontdimen3\font minus
  \fontdimen4\font\relax}
\providecommand{\BIBforeignlanguage}[2]{{%
\expandafter\ifx\csname l@#1\endcsname\relax
\typeout{** WARNING: IEEEtran.bst: No hyphenation pattern has been}%
\typeout{** loaded for the language `#1'. Using the pattern for}%
\typeout{** the default language instead.}%
\else
\language=\csname l@#1\endcsname
\fi
#2}}
\providecommand{\BIBdecl}{\relax}
\BIBdecl

\bibitem{Arik2017DeepVR}
S.~{\"O}. Arik, M.~Chrzanowski, A.~Coates, G.~F. Diamos, A.~Gibiansky, Y.~Kang,
  X.~Li, J.~Miller, A.~Ng, J.~Raiman, S.~Sengupta, and M.~Shoeybi, ``Deep
  voice: Real-time neural text-to-speech,'' in \emph{International Conference
  on Machine Learning}, 2017.

\bibitem{Ren2019FastSpeechFR}
Y.~Ren, Y.~Ruan, X.~Tan, T.~Qin, S.~Zhao, Z.~Zhao, and T.-Y. Liu, ``Fastspeech:
  Fast, robust and controllable text to speech,'' \emph{ArXiv}, vol.
  abs/1905.09263, 2019.

\bibitem{Shen2017NaturalTS}
J.~Shen, R.~Pang, R.~J. Weiss, M.~Schuster, N.~Jaitly, Z.~Yang, Z.~Chen,
  Y.~Zhang, Y.~Wang, R.~J. Skerry-Ryan, R.~A. Saurous, Y.~Agiomyrgiannakis, and
  Y.~Wu, ``Natural tts synthesis by conditioning wavenet on mel spectrogram
  predictions,'' \emph{2018 IEEE International Conference on Acoustics, Speech
  and Signal Processing (ICASSP)}, pp. 4779--4783, 2017.

\bibitem{Ping2017DeepV3}
W.~Ping, K.~Peng, A.~Gibiansky, S.~{\"O}. Arik, A.~Kannan, S.~Narang,
  J.~Raiman, and J.~Miller, ``Deep voice 3: 2000-speaker neural
  text-to-speech,'' \emph{ArXiv}, vol. abs/1710.07654, 2017.

\bibitem{Kumar2021NormalizationDZ}
N.~Kumar, S.~Goel, A.~Narang, and B.~Lall, ``Normalization driven zero-shot
  multi-speaker speech synthesis,'' in \emph{Interspeech}, 2021.

\bibitem{Sun2019TokenLevelED}
H.~Sun, X.~Tan, J.-W. Gan, H.~Liu, S.~Zhao, T.~Qin, and T.-Y. Liu,
  ``Token-level ensemble distillation for grapheme-to-phoneme conversion,''
  \emph{ArXiv}, vol. abs/1904.03446, 2019.

\bibitem{Guo2022UnsupervisedWP}
Y.~Guo, C.~Du, and K.~Yu, ``Unsupervised word-level prosody tagging for
  controllable speech synthesis,'' \emph{ICASSP 2022 - 2022 IEEE International
  Conference on Acoustics, Speech and Signal Processing (ICASSP)}, pp.
  7597--7601, 2022.

\bibitem{Bae2020SpeakingSC}
J.~Bae, H.~Bae, Y.-S. Joo, J.~Lee, G.-H. Lee, and H.-Y. Cho, ``Speaking speed
  control of end-to-end speech synthesis using sentence-level conditioning,''
  \emph{ArXiv}, vol. abs/2007.15281, 2020.

\bibitem{Guo2022PromptTTSCT}
Z.~Guo, Y.~Leng, Y.~Wu, S.~Zhao, and X.~Tan, ``Prompttts: Controllable
  text-to-speech with text descriptions,'' \emph{ArXiv}, vol. abs/2211.12171,
  2022.

\bibitem{Huang2022FastDiffAF}
R.~Huang, M.~W.~Y. Lam, J.~Wang, D.~Su, D.~Yu, Y.~Ren, and Z.~Zhao, ``Fastdiff:
  A fast conditional diffusion model for high-quality speech synthesis,'' in
  \emph{International Joint Conference on Artificial Intelligence}, 2022.

\bibitem{Liu2022DiffGANTTSHA}
S.~Liu, D.~Su, and D.~Yu, ``Diffgan-tts: High-fidelity and efficient
  text-to-speech with denoising diffusion gans,'' \emph{ArXiv}, vol.
  abs/2201.11972, 2022.

\bibitem{Xiao2021TacklingTG}
Z.~Xiao, K.~Kreis, and A.~Vahdat, ``Tackling the generative learning trilemma
  with denoising diffusion gans,'' \emph{ArXiv}, vol. abs/2112.07804, 2021.

\bibitem{Vaswani2017AttentionIA}
A.~Vaswani, N.~M. Shazeer, N.~Parmar, J.~Uszkoreit, L.~Jones, A.~N. Gomez,
  L.~Kaiser, and I.~Polosukhin, ``Attention is all you need,'' \emph{ArXiv},
  vol. abs/1706.03762, 2017.

\bibitem{Devlin2019BERTPO}
J.~Devlin, M.-W. Chang, K.~Lee, and K.~Toutanova, ``Bert: Pre-training of deep
  bidirectional transformers for language understanding,'' \emph{ArXiv}, vol.
  abs/1810.04805, 2019.

\bibitem{Panayotov2015LibrispeechAA}
V.~Panayotov, G.~Chen, D.~Povey, and S.~Khudanpur, ``Librispeech: An asr corpus
  based on public domain audio books,'' \emph{2015 IEEE International
  Conference on Acoustics, Speech and Signal Processing (ICASSP)}, pp.
  5206--5210, 2015.

\bibitem{Ren2020FastSpeech2F}
Y.~Ren, C.~Hu, X.~Tan, T.~Qin, S.~Zhao, Z.~Zhao, and T.-Y. Liu, ``Fastspeech 2:
  Fast and high-quality end-to-end text to speech,'' \emph{ArXiv}, vol.
  abs/2006.04558, 2020.

\bibitem{Kong2020HiFiGANGA}
J.~Kong, J.~Kim, and J.~Bae, ``Hifi-gan: Generative adversarial networks for
  efficient and high fidelity speech synthesis,'' \emph{ArXiv}, vol.
  abs/2010.05646, 2020.

\bibitem{Yang2020VocGANAH}
J.~Yang, J.~Lee, Y.-I. Kim, H.~Cho, and I.~Kim, ``Vocgan: A high-fidelity
  real-time vocoder with a hierarchically-nested adversarial network,''
  \emph{ArXiv}, vol. abs/2007.15256, 2020.

\bibitem{Mao2016LeastSG}
X.~Mao, Q.~Li, H.~Xie, R.~Y.~K. Lau, Z.~Wang, and S.~P. Smolley, ``Least
  squares generative adversarial networks,'' \emph{2017 IEEE International
  Conference on Computer Vision (ICCV)}, pp. 2813--2821, 2016.

\bibitem{Larsen2015AutoencodingBP}
A.~B.~L. Larsen, S.~K. S{\o}nderby, H.~Larochelle, and O.~Winther,
  ``Autoencoding beyond pixels using a learned similarity metric,''
  \emph{ArXiv}, vol. abs/1512.09300, 2015.

\bibitem{DeepSpeech2}
D.~Amodei, S.~Ananthanarayanan, R.~Anubhai, J.~Bai, E.~Battenberg, C.~Case,
  J.~Casper, B.~Catanzaro, Q.~Cheng, G.~Chen, J.~Chen, J.~Chen, Z.~Chen,
  M.~Chrzanowski, A.~Coates, G.~Diamos, K.~Ding, N.~Du, E.~Elsen, and Z.~Zhu,
  ``Deep speech 2: End-to-end speech recognition in english and mandarin,'' 12
  2015.

\bibitem{tacotron}
J.~Shen, R.~Pang, R.~Weiss, M.~Schuster, N.~Jaitly, Z.~Yang, Z.~Chen, Y.~Zhang,
  Y.~Wang, R.~Skerrv-Ryan, R.~Saurous, Y.~Agiomvrgiannakis, and Y.~Wu,
  ``Natural tts synthesis by conditioning wavenet on mel spectrogram
  predictions,'' 04 2018, pp. 4779--4783.

\bibitem{g2p}
G2P, ``G2p, https://github.com/kyubyong/g2,'' 10 2017.

\bibitem{mfa}
M.~McAuliffe, M.~Socolof, S.~Mihuc, M.~Wagner, and M.~Sonderegger, ``Montreal
  forced aligner: Trainable text-speech alignment using kaldi,'' in
  \emph{INTERSPEECH}, 2017.

\bibitem{Chu2001AnOM}
M.~Chu and H.~Peng, ``An objective measure for estimating mos of synthesized
  speech,'' in \emph{Interspeech}, 2001.

\end{thebibliography}
\end{document}